\documentstyle[preprint,eqsecnum,aps,epsfig]{revtex}
\begin{document}
\title{
{\footnotesize {\hfill gr-qc/0305050, DSF 2003/14}}\\
Spacetime noncommutativity and antisymmetric tensor
dynamics in the early universe}
\author{Elisabetta Di Grezia,$^{2,1}$
\thanks{Electronic address: digrezia@na.infn.it}
Giampiero Esposito,$^{1,2}$
\thanks{Electronic address: giampiero.esposito@na.infn.it}
Agostino Funel,$^{2}$
\thanks{Electronic address: funelago@virgilio.it}
Gianpiero Mangano,$^{1,2}$
\thanks{Electronic address: mangano@na.infn.it}
Gennaro Miele$^{2,1}$
\thanks{Electronic address: gennaro.miele@na.infn.it}}
\address{${ }^{1}$Istituto Nazionale di
Fisica Nucleare, Sezione di Napoli,\\
Complesso Universitario di Monte S. Angelo, Via Cintia,
Edificio N', 80126 Napoli, Italy\\
${ }^{2}$Dipartimento di Scienze Fisiche,
Complesso Universitario di Monte S. Angelo,\\
Via Cintia, Edificio N', 80126 Napoli, Italy}
\maketitle
\begin{abstract}
This paper investigates the possible cosmological implications of the
presence of an antisymmetric tensor field $\theta$ related to a
lack of commutativity of spacetime coordinates at the Planck era.
For this purpose, $\theta$ is promoted to a dynamical variable, 
inspired by tensor formalism.
By working to quadratic order in $\theta$, 
we study the field equations in a Bianchi I
universe in two models: an antisymmetric tensor plus scalar
field coupled to gravity, or a cosmological constant and a free
massless antisymmetric tensor. In the first scenario, numerical
integration shows that, in the very early universe, the effects of
the antisymmetric tensor can prevail on the scalar field, while at
late times the former approaches zero and the latter drives the
isotropization of the universe. In the second model, an
approximate solution is obtained of a 
nonlinear ordinary differential equation
which shows how the mean Hubble parameter and the difference between 
longitudinal and orthogonal Hubble parameter evolve in the
early universe.
\end{abstract}


\section{Introduction}

Spacetime noncommutativity is one of the key new hints which
follow from recent developments in quantum field theory. It
has been recently realized, in particular, 
that a consequence of string theory \cite{[1],[2]} is
that the structure of spacetime becomes noncommutative
\cite{[3]}, which can be described loosely as an analog of a
quantum phase space, in terms of the algebra generated by
noncommuting coordinates $[x^{\mu },x^{\nu }]=i\theta^{\mu \nu}$
with $\theta^{\mu \nu }$ an antisymmetric
tensor\footnote{Hereafter we use the natural units $\hbar=c=1$.}.
The idea behind spacetime noncommutativity is very much inspired
by quantum mechanics. A quantum spacetime is defined by replacing
canonical variables with self-adjoint operators
which obey Heisenberg-like commutation relations \(
[x^{\mu},x^{\nu}]=i \theta ^{\mu \nu} \), and can be viewed
as the smearing out of a classical manifold, with the notion of a
point replaced with that of a {\it Planck cell}. It was von
Neumann who first attempted to rigorously describe such a quantum
``space'' and he called this study ``pointless geometry'',
referring to the fact that the notion of a point in a quantum
phase space is meaningless because of the Heisenberg uncertainty
principle of quantum mechanics. This led to the theory of von
Neumann algebras and was essentially the birth of {\it
noncommutative geometry}, referring to the study of topological
spaces whose commutative \( C^{*} \)-algebras of functions are
replaced by noncommutative algebras
\cite{[3],[4],[5],[6],[7],[8]}. The idea of noncommutative
geometry was revived in the eighties by Connes \cite{[5]} and
others, who generalized the notion of a differential structure to
the noncommutative setting, i.e. to arbitrary \( C^{*}
\)-algebras. A theory on a noncommutative space replaces the
noncommutativity of operators associated to spacetime
coordinates with a deformation of the algebra of functions defined
on spacetime. In this context classical general relativity would
break down at the Planck scale because spacetime would no longer
be described by a differentiable manifold, and at these length
scales quantum gravitational fluctuations become large and cannot
be ignored. We stress, however, that
the form of noncommutative geometry we
are interested in is not directly related to current string
theories (see below).

In the past few years several authors \cite{[9],[10],[11]}, 
including some of the
present authors \cite{[12],[13],[14],[15],[16]}, 
have considered the possible effects of noncommutative
geometry and Planck scale physics in cosmology. In
particular, it has been shown that deformation of spacetime and/or
phase space algebras may lead to several interesting features in
the power spectrum of primordial perturbations produced during the
inflationary era \cite{[17],[18],[19],[20]}.
In all these investigations, however,
$\theta^{\mu \nu}$ has been taken to be constant, or with an
a-priori modelled time evolution. In view of general covariance
one may expect that $\theta^{\mu\nu}$ should be rather considered
as a dynamical tensor, coupled to gravity and possibly affecting
the cosmological evolution of the early universe. This is actually 
a crucial point which deserves a thorough treatment, here summarized
by relying in part upon Ref. \cite{[16]}. On the one hand, it is true
that, if one looks at the interplay between string theory and
noncommutative geometry, one has to consider a constant 
$B$-field and hence a constant $\theta^{\mu \nu}$ \cite{[21]}.
On the other hand, the commutator of the $x^{\mu}$ is a tensor,
whose transformation under boosts yields nonvanishing spacetime
components. Moreover, possible violations of unitarity are taken
care of by imposing the conditions \cite{[16]}
$$
\theta^{\mu \nu}\theta_{\mu \nu}>0 \; \; \; , \; \; \;
\varepsilon_{\mu \nu \rho \sigma}
\theta^{\mu \nu}\theta^{\rho \sigma}=0 \; \; \; .
$$
From the point of view of general formalism, it is therefore
legitimate to address the question of whether a broader 
noncommutative picture can be consistently built. This implies
a departure from current models which only exploit a constant
$\theta^{\mu \nu}$, and suggests starting from a
nonlocal action functional with $*$ product of fields in the
presence of nonvanishing spacetime curvature \cite{[16]}.
We therefore assume, hereafter, that the background geometry 
remains a classical pseudo-Riemannian geometry endowed with a
Levi-Civita connection $\nabla$, while the $*$ product of scalar
fields at the same spacetime point is defined by \cite{[16]}
\begin{equation}
\varphi(x) * \psi(x) \equiv \sum_{k=0}^{\infty}{1\over k!}
(i/2)^{k}\theta^{\mu_{1}\nu_{1}}(x)...\theta^{\mu_{k}\nu_{k}}(x)
(\nabla_{\mu_{1}}...\nabla_{\mu_{k}}\varphi)
(\nabla_{\nu_{1}}...\nabla_{\nu_{k}}\psi) \; \; \; .
\label{(1.1)}
\end{equation}
Similarly, having treated classically the geometry, we assume for
tensor fields that
\begin{equation}
F_{\lambda_{1}..\lambda_{s}}*F^{\lambda_{1}...\lambda_{s}}
\equiv g^{\lambda_{1}r_{1}}...g^{\lambda_{s}r_{s}}
F_{\lambda_{1}...\lambda_{s}}*F_{r_{1}...r_{s}} \; \; \; ,
\label{(1.2)}
\end{equation}
where 
\begin{equation}
F_{\lambda_{1}...\lambda_{s}}*F_{r_{1}...r_{s}}
\equiv \sum_{k=0}^{\infty}{1\over k!}(i/2)^{k}
\theta^{\mu_{1}\nu_{1}}(x)...\theta^{\mu_{k}\nu_{k}}(x)
(\nabla_{\mu_{1}}...\nabla_{\mu_{k}}
F_{\lambda_{1}...\lambda_{s}})
(\nabla_{\nu_{1}}...\nabla_{\nu_{k}}
F_{r_{1}...r_{s}}) \; \; .
\label{(1.3)}
\end{equation}
As is stressed in Ref. \cite{[16]}, the occurrence of covariant
derivatives in our definitions (1.1) and (1.3) spoils associativity
of the $*$ product. However, noncommutative effects are already
present at quadratic order in $\theta^{\mu \nu}$, and our
$\theta^{\mu \nu}$ will be taken to be sufficiently small so that 
higher order terms in the action functional are negligible.
 
Another important goal to be pursued is a fully
consistent study of primordial perturbations, which should take into
account the $\theta^{\mu\nu}$ perturbations as well, generalizing
the formalism of gauge-invariant perturbations \cite{[22]} to
anisotropic background metrics.

As a first step in this programme, in this paper we discuss the possible
dynamical evolution of a background, time-dependent antisymmetric tensor
\cite{[23],[24],[25],[26],[27],[28],[29],[30]}
\cite{[31],[32],[33],[34],[35],[36],[37],[38]} in two possible scenarios: a
free massless $\theta^{\mu \nu }$ in presence of a cosmological constant,
the latter being introduced as the easiest way to trigger an inflationary
dynamics, and a more general scenario where the noncommutative field
$\theta^{\mu \nu}$ is coupled to a scalar inflaton. As mentioned the
presence of $\theta^{\mu \nu }$ breaks the isotropy of the universe, and
hence only spatial homogeneity is preserved, leading in turn to a
dependence of $\theta^{\mu \nu }$ on time only. In this framework the
appropriate geometry for the universe is therefore a Bianchi I model. In
Sec. II we obtain a nonlinear system of the background equations in the
presence of inflaton plus a coupling term between inflaton and $\theta^{\mu
\nu}$, and the energy-momentum tensor. In Sec. III we consider the simpler
cosmological term model which however can be worked out analytically, at
least for those initial conditions which are of some interest and may lead
to an early stage where the energy-momentum tensor is dominated by
$\theta^{\mu\nu}$. Concluding remarks and open problems are presented in
Sec. IV, while relevant details are given in the Appendix.

\section{Antisymmetric tensor plus scalar field coupled to Gravity}

In order to describe a field dynamics which might lead to
anisotropy in the early universe, the appropriate model is a
Bianchi I universe (as we stated before) if one wants to preserve
the spatial homogeneity. In this case the line element can be
written as
\begin{equation}
d s^2 = dt^2 - \sum_{i=1}^{3} a_i^2(t) \, (dx^{i})^2 \,\,\, ,
\end{equation}
and correspondingly the nonvanishing connection coefficents are
(no summation over $i$ is here meant)
\begin{equation}
\Gamma_{ij}^0= \delta_{ij}\dot a_i a_i \; \; \; ,~~~~~~
\Gamma_{~0i}^i= \frac{\dot a_i}{a_i} \; \; \; , \; \; \;
\forall \; i,j=1,2,3 \; \; \; ,
\end{equation}
and for the Ricci tensor one has
\begin{equation}
R_{0}^0= -\sum_{i=1}^{3} 
\frac{\ddot a_i}{a_i}\,\,\, ,~~~~~~R_{i}^j= - \delta_i^j
\left(
\frac{\ddot a_i }{a_i}+\frac{\dot a_i}{a_i}
\sum_{k\neq i} \frac{\dot a_k}{a_k}
\right)\,\,\, .
\end{equation}
We consider a model in which there are both the antisymmetric
tensor responsible of noncommutativity of spacetime and a
minimally coupled massive scalar field which drives the inflation. 
The corresponding nonlocal action reads 
\begin{eqnarray}
S &=& \int d^{4}x \sqrt{-g} \left[-{R\over 16 \pi G}
+{1\over 12}H_{\mu \nu \sigma}*H^{\mu \nu \sigma}
+{1\over 2}\varphi_{;\mu}*\varphi^{;\mu}
-{m^{2}\over 2}\varphi* \varphi \right .\nonumber \\
&-& \left . {\lambda \over 2} (\varphi * \varphi)
* (\theta_{\mu \nu} * \theta^{\mu \nu}) \right] \; \; \; \; ,
\label{(2.4)}
\end{eqnarray}
where the part involving $\lambda$ is here introduced to mimic
a `time-dependent' mass term for $\theta^{\mu \nu}$, and
\begin{equation}
H_{\mu \nu \sigma}  \equiv \nabla _{\mu }\theta _{\nu \sigma }+
\nabla _{\nu }\theta _{\sigma \mu }
+\nabla_{\sigma }\theta _{\mu \nu }
\label{(2.5)}
\end{equation}
is the field strength associated to the antisymmetric tensor \(
\theta_{\mu \nu }= -\theta_{\nu \mu }\) (hereafter, Greek indices
run from $0$ through $3$, whereas Latin indices run from $1$
through $3$). It should be noticed
that the kinetic term for $\theta _{\mu \nu }$ 
is inspired by a generalization of the Maxwell
theory \cite{[34]} and that in Eq.(\ref{(2.5)}) only the effects of
partial derivatives survive \footnote{Our notation agrees with the
one used, for example, in Ref. \cite{[35]}.}.
Now, by virtue of the definitions (1.1)--(1.3), one finds
\begin{equation}
H_{\mu \nu \sigma}* H^{\mu \nu \sigma}
=H_{\mu \nu \sigma}H^{\mu \nu \sigma}
+{\rm O}(\theta^{3}) \; \; \; ,
\label{(2.6)}
\end{equation}
\begin{equation}
\varphi * \varphi=\varphi^{2}-{1\over 8}\theta^{\mu \nu}
\theta^{\rho \sigma}(\nabla_{\mu}\nabla_{\rho}\varphi)
(\nabla_{\nu}\nabla_{\sigma}\varphi)
+{\rm O}(\theta^{3}) \; \; \; ,
\label{(2.7)}
\end{equation}
\begin{equation}
(\varphi * \varphi) * (\theta_{\mu \nu} * \theta^{\mu \nu})
=\varphi^{2}\theta_{\mu \nu}\theta^{\mu \nu}
+{\rm O}(\theta^{3}) \; \; \; .
\label{(2.8)}
\end{equation}
Thus, to second order in $\theta^{\mu \nu}$, which is relevant
since $\theta^{\mu \nu}$ is taken to be sufficiently small, only
kinetic and potential term for the scalar field contribute, but 
with vanishing coefficient, since the former changes by the 
amount \cite{[16]}
(integration by parts yields also a third term which however vanishes 
if $\varphi=\varphi(t)$ only)
\begin{equation}
\delta S_{K}={1\over 32}\int d^{4}x \sqrt{-g} \; \theta^{\mu \nu}
\theta^{\rho \sigma}(\nabla_{\rho}\nabla_{\tau}\varphi)
([\nabla_{\mu},\nabla_{\nu}]\nabla_{\sigma}
\nabla^{\tau}\varphi) \; \; \; ,
\label{(2.9)}
\end{equation}
and the latter changes by the amount \cite{[16]}
\begin{equation}
\delta S_{m}={m^{2}\over 32}\int d^{4}x \sqrt{-g} \;
\theta^{\mu \nu}\theta^{\rho \sigma}
R_{\; \sigma \mu \nu}^{\tau}
(\partial_{\tau}\varphi)(\partial_{\rho}\varphi) \; \; \; .
\label{(2.10)}
\end{equation}
Since $\varphi$ depends only on the time variable, both (2.9) and
(2.10) vanish in our Bianchi I background. Thus, to quadratic
order in $\theta^{\mu \nu}$, we end up with the local action
functional
\begin{equation}
S \equiv \int  d^{4} x \sqrt{-g} \textrm{ }\left[-\frac{R}{16\pi
G}+\frac{1}{12}
\textrm{H}_{\mu \nu \sigma}\textrm{H} ^{\mu \nu \sigma}
+\frac{1}{2} \varphi _{;\mu }\varphi ^{;\mu }-V(\varphi )-\frac{\lambda
}{2} \, \varphi^{2} \,
\theta _{\mu \nu } \theta ^{\mu \nu }\right] \,\,\,,
\label{action}
\end{equation}
where $V(\varphi) \equiv {m^{2}\over 2}\varphi^{2}$ hereafter.
At this stage, the resulting energy-momentum tensor is given by
\begin{eqnarray}
T^{\beta }_{\alpha }&=&\delta ^{\beta }_{\alpha }\left[-\frac{1}{12}
\textrm{H}_{\mu \nu \sigma }\textrm{H}^{\mu \nu \sigma}
-\frac{1}{2}\varphi _{;\mu }\varphi ^{;\mu }+V(\varphi )
+\frac{\lambda}{2} \, \varphi ^{2} \, \theta _{\mu \nu }\theta ^{\mu \nu
}\right]
\nonumber\\
&+&\frac{1}{2}\textrm{H}_{\alpha \mu \nu }\textrm{H}^{\beta \mu \nu
}+\varphi _{;
\alpha }\varphi ^{;\beta }-2\, \lambda \, \varphi ^{2}\theta _{\alpha\nu }
\theta ^{\beta \nu }\,\,\,.
\label{2a}
\end{eqnarray}
By using the expressions (\ref{action}), (\ref{2a}) the resulting
equations of motion are
\begin{eqnarray}
&&\bigtriangledown^{\mu }\textrm{H}_{\mu \nu \sigma }+2 \, \lambda
\,
\varphi
^{2}\theta
_{\nu \sigma }=0 \,\,\, , \label{eq:theta}
\\
&&\bigtriangledown _{\mu }\bigtriangledown ^{\mu }\varphi + \lambda
\, \varphi
\,
\theta
_{\mu \nu }\theta ^{\mu \nu }+\frac{\delta V}{\delta \varphi}=0 \,\,\, ,
\label{eq:phi}
\\
&&R_{\mu \nu }-\frac{1}{2} \, g_{\mu \nu }\, R \, =\, 8 \, \pi \, G \,
T_{\mu
\nu }
\,\,\,
.
\label{eq:metric}
\end{eqnarray}
To be consistent with the spacetime homogeneity ansatz
we assume that all fields are depending on time only. In this
case the above equations read (hereafter $H_i \equiv
\dot{a}_i/a_i$)
\begin{eqnarray}
&&\lambda \, \varphi^2 \, \theta_{0i} = 0 \,\,\, , \label{eq:theta0i}\\
&&\ddot{\theta}_{ij} + \left(\sum_{k=1}^3 H_k \right) \dot{\theta}_{ij} - 2
\left( H_i + H_j \right) \dot{\theta}_{ij}
+ 2 \, \lambda \, \varphi^2 \, \theta_{ij}
=0 \,\,\, , \label{eq:thetaij}\\
&&\ddot \varphi + \left(\sum_{k=1}^3 H_k \right) \dot \varphi  +
\lambda \, \varphi \, \theta^{\mu \nu}\theta_{\mu \nu} +
\frac{\delta V}{ \delta \varphi} = 0 \,\,\, , \label{eq:varphi1}\\
&& \sum_{k=1}^3 \left( \dot{\theta}_{ik} \, \dot{\theta}_{jk} - 2 \,
\lambda
\,\varphi^2 \, \theta_{ik} \, \theta_{jk} \right) =0
\,\,\, , \,\,\,\,\,\,\,\,
\forall i \neq j \label{eq:rij}\\
&&\sum_i \frac{\ddot a_i}{a_i} = 8 \, \pi \, G \,\left( V(\varphi)
- \dot \varphi^2 - \frac{\lambda}{2} \, \varphi ^{2} \,
\theta _{\mu \nu }\theta ^{\mu \nu } \right) \,\,\, , \label{eq:r00}\\
 &&\frac{\ddot a_i
}{a_i}+\frac{\dot a_i}{a_i} \sum_{k\neq i} \frac{\dot a_k}{a_k} =
8 \, \pi \, G \, \left( \frac{1}{6} \textrm{H}_{\mu \nu \sigma
}\textrm{H}^{\mu \nu \sigma} + V(\varphi) -\frac{\lambda}{2} \,
\varphi ^{2} \, \theta _{\mu \nu }\theta ^{\mu \nu } - \frac{1}{2}
\textrm{H}_{i \nu \sigma }\textrm{H}^{i \nu \sigma} + 2  \,
\lambda \, \varphi ^{2} \, \theta _{i \nu }\theta^{i \nu }
\right)\,\,\, .\label{eq:rii}
\end{eqnarray}
Of course, Eqs. (\ref{eq:theta0i}) and (\ref{eq:thetaij}) result
from (\ref{eq:theta}), whereas (\ref{eq:varphi1}) is obtained from
(\ref{eq:phi}). The remaining equations are the Einstein equations
where, in particular, (\ref{eq:rij}) provides a Bianchi I
universe.

Using Eq. (\ref{eq:theta0i}) one easily gets $\theta ^{0i}=\theta
_{0i}=0$. By virtue of Eq. (\ref{eq:rij}) one can show that the
only possible solution has only one nonvanishing component of
$\theta_{ij}$, e.g. $\theta_{12}$. Moreover, since
isotropy is broken and the residual invariance is $SO(2)$, it is
rather natural to choose $a_{1}=a_{2} \equiv a_{\perp}$, $a_{3}
\equiv a_{L}$, with corresponding Hubble parameters $H_{\perp}
\equiv {\dot a}_{\perp}/a_{\perp}$, $H_{L} \equiv {\dot
a}_{L}/a_{L}$. In this case the equations become
\begin{eqnarray}
&&\ddot{\theta}_{12} + \left( H_L - 2 H_\perp \right)
\dot{\theta}_{12} + 2 \, \lambda \, \varphi^2 \, \theta_{12}
=0 \,\,\, , \label{eq:theta12}\\
&&\ddot \varphi + \left(H_L + 2 H_\perp \right) \dot \varphi  +
\lambda \frac{\varphi \, \theta _{12 }^2} {a_\perp^4} +
\frac{\delta V}{ \delta \varphi} = 0 \,\,\, , \label{eq:varphi}\\
&& H^2_\perp + 2 H_\perp H_L = 8 \, \pi \, G \,\left( \frac 12 \,
\dot{\varphi}^2 + V(\varphi)+ \frac 12 \, \frac{\dot{\theta} _{12
}^2} {a_\perp^4} + \lambda \, \frac{\varphi ^{2} \, \theta _{12
}^2} {a_\perp^4} \right) \,\,\, , \label{eq:r0012}\\
&&\dot{H}_\perp + H_\perp \left(H_L + 2 H_\perp \right) = 8 \, \pi
\, G \, \left( V(\varphi) + \, \lambda \, \frac{\varphi ^{2} \,
\theta _{12 }^2} {a_\perp^4} \right)\,\,\, , \label{eq:hperp}\\
&&\dot{H}_L + H_L \left(H_L + 2 H_\perp \right) = 8 \, \pi \, G \,
\left( \frac{\dot{\theta} _{12 }^2} {a_\perp^4} + V(\varphi) - \,
\lambda \, \frac{\varphi ^{2} \, \theta _{12 }^2} {a_\perp^4}
\right)\,\,\, . \label{eq:hlong}
\end{eqnarray}
Let us define the mass parameter $\mu \equiv m_{Pl}/\sqrt{8 \pi}=1/\sqrt{8
\pi G}$. In terms of this quantity we can write a dimensionless system of
differential equations. By defining
$$
x \equiv \varphi/\mu, \;
y \equiv \theta _{12 }/(\mu a_\perp^2), \;
\widetilde{H}_\perp \equiv
H_\perp/\mu, \;
\widetilde{H}_L \equiv H_L/\mu, \;
\widetilde{V}(x)
\equiv V(\mu x)/\mu^4,
$$
and using the dimensionless time $\tau
\equiv t \mu$ we get
\begin{eqnarray}
&&y'' + \left( \widetilde{H}_L + 2 \widetilde{H}_\perp \right) y'
+ 2 \left(\lambda \, x^2 - 2 \, \widetilde{H}^2_\perp +
\widetilde{V}(x) + \lambda \, x^2 \, y^2 \right) y =0
\,\,\, , \label{pi}
\\
&&x'' + \left(\widetilde{H}_L + 2 \widetilde{H}_\perp \right) x' +
\lambda \, x \, y^2 +  \frac{\delta \widetilde{V}}{ \delta x} = 0
\,\,\, ,\label{pio}
\\
&& \widetilde{H}^2_\perp + 2 \widetilde{H}_\perp \widetilde{H}_L =
\frac{{x'}^2}{2} \,
 + \widetilde{V}(x)+ \frac 12 \,
\left(y' + 2 \widetilde{H}_\perp \, y \right)^2  \,
  + \lambda \, x^{2} \, y^2 \,\,\, , \\
&&\widetilde{H}_\perp' + \widetilde{H}_\perp \left(\widetilde{H}_L
+ 2 \widetilde{H}_\perp \right) =  \widetilde{V}(x) + \, \lambda
\, x^2 \, y^2 \,\,\, , \label{pii}\\
&&\widetilde{H}_L' + \widetilde{H}_L \left(\widetilde{H}_L + 2
\widetilde{H}_\perp \right) = \left(y' + 2 \widetilde{H}_\perp \,
y \right)^2 + \widetilde{V}(x) - \, \lambda \, x^2 \, y^2 \,\,\, ,
\end{eqnarray}
where the `prime' denotes the derivative with respect to $\tau$.
The truly independent equations are given by (\ref{pi}),
(\ref{pio}), (\ref{pii}) jointly with
\begin{equation}
 \left( \widetilde{H}_L + 2 \widetilde{H}_\perp \right) =
\frac{1}{2 \,\widetilde{H}_\perp } \left(\frac{{x'}^2}{2} \,
 + \widetilde{V}(x)+ \frac 12 \, \left(y'
+ 2 \widetilde{H}_\perp \, y \right)^2  \,
  + \lambda \, x^{2} \, y^2 + 3 \, \widetilde{H}_\perp^2\right) \,\,\, .
\end{equation}
On recalling the e-folding definition in the orthogonal direction,
i.e. $N_{\perp}(\tau)\equiv 
\log(a_\perp(\tau)/a_\perp(\tau_i))$, one can
easily prove that
\begin{eqnarray}
\frac{d}{d \tau} &=& \widetilde{H}_\perp \, 
\frac{d}{d N_{\perp}} \; \; \; , \\
\frac{d^2}{d \tau^2} &=& \widetilde{H}_\perp^2 \, \frac{d^2}{d
N_{\perp}^2} \, + \, \widetilde{H}_\perp' \, 
\frac{d}{d N_{\perp}} \; \; \; .
\end{eqnarray}
By using $N_{\perp}$ as the evolution parameter and defining $z \equiv
\widetilde{H}_\perp^2 $ we find
\begin{eqnarray}
&&z \, \frac{d^2 \, y}{d N_{\perp}^2} + \left(\widetilde{V}(x) + \,
\lambda \, x^2 \, y^2 \right) \frac{d \, y }{d N_{\perp}}   + 2
\left(\lambda \, x^2 - 2 \, z + \widetilde{V}(x) + \lambda \, x^2
\, y^2 \right) y =0
\,\,\, , \\
&&z \, \frac{d^2 \, x}{d N_{\perp}^2}+ \left(\widetilde{V}(x) + \, \lambda
\, x^2 \, y^2 \right) \frac{d \, x }{d N_{\perp}} 
+ \lambda \, x \, y^2 +
\frac{\delta \widetilde{V}}{ \delta x} = 0
\,\,\, ,\\
&& \frac{d \, z}{d N_{\perp}}  = \widetilde{V}(x) + \lambda \,
x^2 \, y^2 - \frac 12 \, z \, \left[ \left( \frac{d \,x}{d
N_{\perp}}\right)^2 
+ \left( \frac{d \, y}{d N_{\perp}} + 2 y \right)^2 + 6
\right].
\end{eqnarray}
In the following figures we
show, on choosing different initial conditions, that a range of
$N_{\perp}$ exists in which the $\theta_{12}$
component of the $\theta$ field dominates on the
$\varphi$ field; of course, this results from a special choice 
of initial conditions (see following section for a more 
thorough discussion of underlying issues).
Moreover, $H_L$ dominates as well, i.e., there is anisotropy.
Outside this range, $\theta_{12}$ approaches zero and
$\varphi$ dominates driving the isotropization,
which can be seen in the figures
where $H_L$ and $H_\perp$ reach the same constant value.

\begin{figure}
\begin{center}
\epsfig{height=4truecm,file=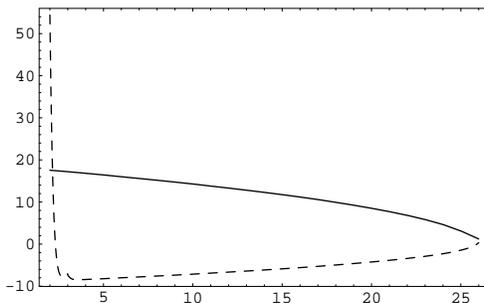}
\caption{The $x$-axis corresponds to the e-folding 
parameter $N_{\perp}$ and on
the $y$-axis we plot $H_{L}$ and $H_{\perp}$ with dashed and
continuous line, respectively; we take
$\widetilde{V}(x)= qx^2$, with initial conditions:
$H_{\perp}^{2} =10, \theta_{12}=100, \frac{d \,
\theta_{12}}{d N_{\perp}}=0, \varphi=10,
\frac{d \,\varphi}{d N_{\perp}}=0$,
and coupling constants $\lambda=10^{-4},q=10$.}
\end{center}
\label{fig1}
\end{figure}
\noindent
\begin{figure}
\begin{center}
\epsfig{height=4truecm,file=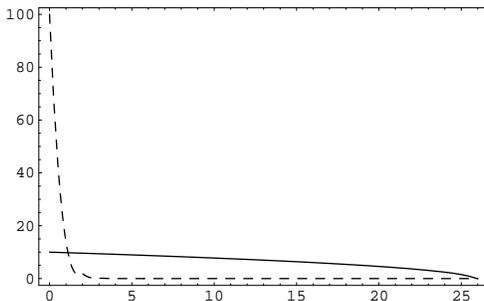}
\caption{The $x$-axis corresponds to the e-folding 
parameter $N_{\perp}$,
and on the $y$-axis we plot $\theta_{12}$ and $\varphi$
with dashed and continuous line, respectively,
with same initial conditions of the previous figure.}
\end{center}
\label{fig11}
\end{figure}
\noindent
\begin{figure}
\begin{center}
\epsfig{height=4truecm,file=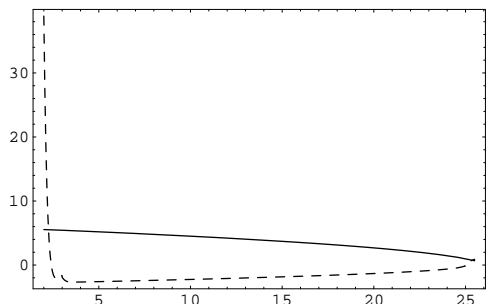}
\caption{We plot $H_{L}$ and $H_{\perp}$ as functions of
the e-folding parameter $N_{\perp}$, with
$\widetilde{V}(x)= qx^2$, and initial conditions:
$H_{\perp}^{2} =10, \theta_{12}=100, \frac{d \,
\theta_{12}}{d N_{\perp}}=1, \varphi=10,
\frac{d \,\varphi}{d N_{\perp}}=0$,
and coupling constants $\lambda=-10^{-5},q=1$.}
\end{center}
\label{fig2}
\end{figure}
\noindent
\begin{figure}
\begin{center}
\epsfig{height=4truecm,file=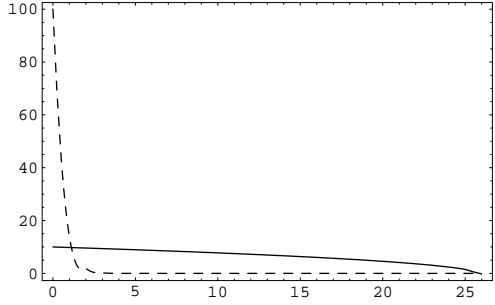}
\caption{The fields $\theta_{12}$ and $\varphi$ are plotted as
functions of the e-folding parameter $N_{\perp}$.}
\end{center}
\label{fig22}
\end{figure}

\section{Cosmological constant and massless antisymmetric tensor}

While the equations in the previous section can only be solved by
means of numerical methods, one can envisage a simpler scenario
which embodies the main features but allows for an analytic
approach. For this purpose we consider a Lagrangian density for
the antisymmetric tensor $\theta_{\mu \nu}$ and gravity including
a cosmological constant term
(with $H_{\mu\nu\rho}$ defined in Eq. (2.5)) 
\begin{eqnarray}
{\cal L}&=& \sqrt{-g} \left[ - \frac{1}{16 \pi G}\, (R + 2 \Lambda)
+\frac{1}{12} \, H_{\mu\nu\rho}*H^{\mu\nu\rho} \right] \nonumber \\
&=& \sqrt{-g}\left[-{1\over 16 \pi G}(R+2\Lambda)
+{1\over 12}H_{\mu \nu \rho}H^{\mu \nu \rho}\right]
+{\rm O}(\theta^{3}) \; \; \; .
\label{(3.1)}
\end{eqnarray}
Thus, on working to quadratic order in $\theta^{\mu \nu}$,
the resulting action is invariant under the gauge transformations
\begin{equation}
\theta_{\mu\nu}\rightarrow \theta_{\mu\nu} + \partial_\mu \chi_\nu
-\partial_\nu\chi_\mu \; \; \; ,
\end{equation}
and the $\theta_{\mu \nu}$ dynamics can be also recast in a
different form by introducing the pseudo-scalar Kalb--Ramond field
$\chi$ via the duality transformation $\partial_\mu \chi =
\epsilon_{\mu\nu\rho\sigma}H^{\nu\rho\sigma}$, with
$\epsilon_{\mu\nu\rho\sigma}$ the volume four-form. Actually,
since we are assuming homogeneity and a purely time-dependent
$\theta$ field this implies a time-independent $\chi$ field with a
linear dependence on spatial coordinates (see Eq. (3.15)).

Using the results of the previous section in the particular case we
are considering, the resulting equations of motion are
\begin{eqnarray}
&&\bigtriangledown^{\mu }\textrm{H}_{\mu \nu \sigma }=0 \,\,\, ,
\label{eq:theta1}
\\
&&R_{\mu \nu }-\frac{1}{2} \, g_{\mu \nu }\, R \, =\, 8 \, \pi \,
G \, T_{\mu \nu }(\theta) + \Lambda \, g_{\mu \nu}\,\,\, ,
\label{eq:metric1}
\end{eqnarray}
and the energy-momentum tensor reads
\begin{eqnarray}
T_\mu^\nu(\theta) = \frac{1}{2}H_{\mu\lambda\rho}
H^{\nu\lambda\rho}-\frac{\delta_\mu^\nu}{12}
H_{\lambda\sigma\rho}H^{\lambda\sigma\rho} \,\,\, .
\end{eqnarray}
As in the previous section, we assume Bianchi I geometry. The
equations of motion (\ref{eq:theta1}) and (\ref{eq:metric1}) now
read
\begin{eqnarray}
&&\ddot{\theta}_{ij} + \left(\sum_{k=1}^3 H_k \right)
\dot{\theta}_{ij} - 2 \left( H_i + H_j \right) \dot{\theta}_{ij}
=0 \,\,\, , \label{eq2:thetaij}\\
&& \sum_{k=1}^3 \dot{\theta}_{ik} \, \dot{\theta}_{jk}  =0 \,\,\,
, \,\,\,\,\,\,\,\,
\forall i \neq j \label{eq2:rij}\\
&&\sum_{i=1}^{3} \frac{\ddot a_i}{a_i} 
= \Lambda \,\,\, , \label{eq2:r00}\\
 &&\frac{\ddot a_i
}{a_i}+\frac{\dot a_i}{a_i} \sum_{k\neq i} \frac{\dot a_k}{a_k} =
\Lambda \, - \, 4 \, \pi \, G \, \left( \textrm{H}_{i \nu \sigma
}\textrm{H}^{i \nu \sigma} \, - \, \frac{1}{3} \, \textrm{H}_{\mu
\nu \sigma }\textrm{H}^{\mu \nu \sigma} \right)\,\,\,
.\label{eq2:rii}
\end{eqnarray}
Unlike the previous section, Eq. (\ref{eq:theta0i}) is now
replaced by a gauge-fixing condition. In the following we will use
the gauge $\theta_{0i}$ = 0. A detailed analysis of the
admissibility of this gauge fixing is reported in the Appendix.

The effect of (\ref{eq2:thetaij}) and (\ref{eq2:rij}) is to restrict
severely the possible choice of initial conditions for $\dot{\theta}_{ij}$.
In particular either all $\dot{\theta}_{ij}(0)$ values vanish and thus
$\theta_{ij}$ is a constant tensor, or at most only one component
$\dot{\theta}_{ij}(0)$ is nonvanishing. Since the antisymmetric tensor
enters the field equations only through its first- and second-order time
derivatives the case of a constant $\theta$-tensor reduces to a pure
$\Lambda$-term cosmology and is thus uninteresting for our analysis. We
therefore consider the case $\dot{\theta}_{13}(0) =\dot{\theta}_{23}(0) =
0$, $\dot{\theta}_{12}(0) \neq 0$, and we eventually get
\begin{eqnarray}
&&\ddot{\theta}_{12} + \left( H_L - 2 H_\perp \right)
\dot{\theta}_{12}
=0 \,\,\, , \label{eq2:theta12}\\
&& H^2_\perp + 2 H_\perp H_L = \Lambda + 4 \, \pi \, G \,
\frac{\dot{\theta} _{12 }^2} {a_\perp^4} \,\,\, , \label{eq2:r0012}\\
&&\dot{H}_\perp + H_\perp \left(H_L + 2 H_\perp \right)
= \Lambda\,\,\, , \label{eq2:hperp}\\
&&\dot{H}_L + H_L \left(H_L + 2 H_\perp \right) = \Lambda + 8 \,
\pi \, G \, \frac{\dot{\theta} _{12 }^2} {a_\perp^4} \,\,\, .
\label{eq2:hlong}
\end{eqnarray}
Equation (\ref{eq2:theta12}) determines the behavior of
$\theta_{12}$ in terms of the scale factors
\begin{equation}
\dot \theta_{12}\, = \, \dot \theta_{12}(0)  \,
\frac{a_\perp^2}{a_L}\,\,\, ,
\label{theta12scale}
\end{equation}
where in (\ref{theta12scale}), by virtue of 
the arbitrariness in the choice of
$a_L(0)$ and $a_\perp(0)$, we have assumed, without loss of generality,
$a_\perp(0)=a_L(0)=1$. Notice that this immediately gives the spacetime
dependence of the Kalb--Ramond scalar field
\begin{equation}
\chi = \dot \theta_{12}(0) \, x^3 + \chi_0 \,\,\, ,
\label{chi}
\end{equation}
where, according to (2.1),  
$x^3$ denotes the longitudinal spatial coordinate. The equation
(\ref{chi}) is actually consistent with spatial homogeneity, while it
breaks isotropy.

The previous equations can be recast in a simpler and more useful
form by introducing the variable
\begin{equation}
\xi \equiv \frac13 \log (a_\perp^2 a_L) \,\,\, , \label{xi}
\end{equation}
which represents the average e-folding. 
Hence one gets
\begin{eqnarray}
H^2_\perp + 2 H_\perp H_L &=& \Lambda
+ \frac{c^2}{2 \, a_L^2} \,\,\, , \label{1}\\
\left( H_\perp' + 3H_\perp \right)
\left( H_L+2 H_\perp \right) &=& 3 \Lambda \,\,\, ,\label{2} \\
\left( H_L' + 3 H_L \right) \left( H_L+2 H_\perp\right) &=& 3
\Lambda + 3 \frac{c^2}{a_L^2} \,\,\, ,\label{3}
\end{eqnarray}
where $c^2 \equiv 8 \pi G \left(\dot \theta_{12}(0)\right)^2$ and
$' \equiv d/d\xi$. It is also convenient to set
\begin{eqnarray}
a_L(\xi) = \exp\left\{ \xi + \frac{\Omega(\xi)}{2} \right\}\,\,\,
, \label{3.18}
\end{eqnarray}
and hence from (\ref{xi})
\begin{eqnarray}
a_\perp(\xi) = \exp\left\{\xi -\frac{\Omega(\xi)}{4}\right\}\,\,\,
. \label{3.19}
\end{eqnarray}
Furthermore, we introduce the mean Hubble parameter, $H \equiv
(H_L+2H_\perp)/3$, and the asymmetry function $h \equiv H_L-H_\perp$. 
Since $a_L(0)=a_\perp(0)=1$, Eqs. (3.20) and (3.21)
yield $\Omega(0)=0$. Thus one gets
\begin{eqnarray}
\frac{1}{2} \left( H^2\right)' \, + \, 3H^2 &=& \Lambda +
\frac{c^2}{3}\,
\exp\left\{ -2 \xi -\Omega(\xi)\right\} \,\,\, , \label{1p}\\
h' + 3h&=&  \frac{c^2}{H} \, \exp\left\{ -2 \xi
-\Omega(\xi)\right\}
\,\,\, ,\label{2p} \\
H^2 - \frac19 \,h^2  &=& {\Lambda \over 3}  + {c^{2}\over 6} \,
\exp\left\{ -2 \xi -\Omega(\xi)\right\}\,\,\, ,\label{3p}
\end{eqnarray}
where the last equation is the Hamiltonian constraint. Actually
the functional dependence of $h$ in terms of $H$ and $\Omega$ can
be already obtained by its very definition
\begin{equation}
h\equiv H \left(\frac{a_L'}{a_L} - \frac{a_\perp'}{a_\perp}
\right) = \frac 34 H \Omega'\,\,\, , \label{hversusomega}
\end{equation}
showing that $\Omega'$ is directly related to the isotropy
breaking. In the following we will consider an expanding universe
in both longitudinal and transverse directions, $H_L$,
$H_\perp>0$. In this case we note that $-2 \leq \Omega' <4$. We
will use this information later on. Notice that $\Omega'=4$ would
correspond to $H_\perp=0$, which is forbidden by (\ref{1}).

The equation (\ref{1p}) is a first-order inhomogeneous equation
whose solution reads
\begin{equation}
H^2(\xi) = \frac{\Lambda}{3} \, + \,
\left(H^2(0)-\frac{\Lambda}{3} \right) \exp\left( - 6 \xi\right)
\, + \frac 23 \, c^2 \, \exp\left( - 6 \xi\right) \, \int^\xi_0
\exp\left( 4 \zeta -
\Omega(\zeta)\right) \, d\zeta \,\,\, ,\label{hT2}\\
\end{equation}
which, together with (\ref{hversusomega}), gives the Hubble
parameters in terms of the function $\Omega(\xi)$. The latter can
be determined as the solution of a second-order differential
equation, obtained by inserting (3.25) into the differential
equation (3.23) and the Hamiltonian constraint (3.24). The latter
operation yields
$$
H^{2}=\left(1-{{\Omega'}^{2}\over 16}\right)^{-1}
\left({\Lambda \over 3}+{c^{2}\over 6}\exp(-2\xi-\Omega(\xi))\right).
$$
If the differential equation (3.22) is also exploited to express
$HH'$, we eventually find
\begin{eqnarray}
\Omega''(\xi) &+& 2 \Omega'(\xi) \left(
1-\frac{\Omega'(\xi)^2}{16} \right)\frac{3 \Lambda + c^2
\exp\left(-2 \xi -\Omega(\xi)\right)}{2 \Lambda+ c^2 \exp\left(-2 \xi
-\Omega(\xi) \right)} \\
\nonumber 
&-& 8 \left( 1-\frac{\Omega'(\xi)^2}{16} \right)\frac{
c^2 \exp\left(-2 \xi -\Omega(\xi) \right)}{2 \Lambda+ c^2
\exp\left(-2 \xi -\Omega(\xi) \right)}=0 \,\,\, , \label{omegaeq}
\end{eqnarray}
with initial conditions $\Omega(0)=0$, $\Omega'(0)=(4/3)
h(0)/H(0)$.

This equation can be hardly solved analytically in the general case, but it
reduces to much simpler forms in the two regimes, when either the
cosmological constant or the $\theta_{\mu\nu}$ fields dominate the
energy-momentum tensor. At very early times the latter is likely to be
largely the dominant component. We can rewrite the parameter $c^2$ as
$c^2=m_{NC}^4/m_{Pl}^2$, with $m_{NC}$ the scale where the classical
picture of spacetime manifold breaks down. The order of magnitude of
$\Lambda$ can be instead constrained by the fact that it drives inflation
in the late stages, and represents in the slow-roll approximation the
potential of the inflaton field. This is severely bounded by the fact that
it should account for the correct amplitude of primordial perturbations.
For example, for a polynomial potential $V(\phi) =\lambda \phi^n/n$, the
requirement of slow-roll dynamics and perturbation amplitudes of the order
of $10^{-5}$ gives  $\Lambda
\leq 10^{-12} m_{Pl}^2$ \cite{[38]}. As long as $m_{NC} \sim
m_{Pl}$ the value of $\Lambda$ is several orders of magnitude
smaller than $c^2$, so in this case the early dynamics is fully
determined by $\theta_{\mu\nu}$. Smaller values of $m_{NC}$, such
that $c^2 \sim \Lambda$, cannot be ruled out of course, but in this
case Eq. (3.27) can only be solved numerically.

Hereafter we specialize to the case $m_{NC} \sim m_{Pl}$. At early
times therefore Eq. (3.27) takes the simplified form
\begin{equation}
\Omega''(\xi) + 2 (\Omega'(\xi)-4) \left(
1-\frac{\Omega'(\xi)^2}{16} \right)=0 \,\,\, . \label{omegaearly}
\end{equation}
Since the $\theta_{\mu\nu}$ contribution is diluted with expansion
as $a_L^2$, this equation holds approximatively for values of
$\xi$ smaller than the value $\xi_*$ such that $\Lambda \sim
c^2/a_{L}^{2}(\xi_*)$, i.e.
\begin{equation}
\xi_* + \frac{\Omega(\xi_*)}{2} \sim \log \frac{c^2}{\Lambda}\,\,\, .
\label{xistar}
\end{equation}
Later expansion is instead driven by $\Lambda$ and hence, by
neglecting the $\theta_{\mu\nu}$ contribution, we have
\begin{equation}
\Omega''(\xi) + 3 \Omega'(\xi) \left( 1-\frac{\Omega'(\xi)^2}{16}
\right)=0 \,\,\, . \label{lateevol}
\end{equation}
This equation of course should give back the isotropization phase
leading to a de Sitter phase.

We begin by studying the early time evolution. Before doing this it
is worth discussing the values of initial conditions for
Hubble parameters. From Eqs. (\ref{eq2:hperp}) and (\ref{eq2:hlong})
we see that $\theta_{\mu\nu}$ acts as source for $a_L$ only, the
evolution of $a_\perp$ being expected to be much slower. In other
words, {\it the antisymmetric tensor drives the expansion of the
longitudinal scale factor only}. The most natural choice at $\xi=0$
is therefore $H_L(0)>>H_\perp(0)$, that is to say
$\Omega'(0)=4-12 \epsilon$, with $\epsilon=H_\perp(0)/H_L(0) <<1$.
In this case the solution of Eq. (\ref{omegaearly}) is particularly
simple. On defining $\lambda(\xi) \equiv \Omega(\xi)-4 \xi$, the latter
reduces to
\begin{equation}
\lambda''(\xi) = {\lambda'}^{2}(\xi) \left( 1+ \frac{\lambda'(\xi)}{8}
\right)\,\,\, .
\end{equation}
Upon considering $\log \lambda'(\xi) \equiv y(\xi)$, this equation is solved
exactly by separation of variables and subsequent integration,
and yields ($C$ being an integration constant)
\begin{equation}
\xi+C=-\exp(-y)-{y\over 8}+{1\over 8}
\log \left(1+{1\over 8}\exp(y)\right) \,\,\, .
\end{equation}
On choosing the initial conditions
\begin{equation}
\lambda(0)=0, \; \lambda'(0)=-12 \epsilon\,\,\, ,
\end{equation}
the approximate solution at small $\epsilon$ reads, with a
very good accuracy,
\begin{equation}
\lambda(\xi) = -\frac{1}{1-{3 \over 2}\, \epsilon} 
\log \left(1+ 12 \, \epsilon
\left(1-\frac{3 }{2}\, \epsilon \right) \xi \right)\,\,\, .
\label{lambdasol}
\end{equation}
For example, we have checked numerically that this solution is
accurate at the per thousand level
up to $\xi =10^{2}$, if $\epsilon<
10^{-2}$. These ranges fully cover the early stage. In fact Eq.
(\ref{omegaearly}) no longer holds at $\xi_*$, see
(\ref{xistar}), which, in view of the logarithmic behavior of
$\lambda(\xi)$, for $\epsilon< 10^{-2}$ is approximatively fixed
by
\begin{equation}
\xi_* \sim \frac{1}{3} \log \frac{c^2}{\Lambda}\,\,\, .
\end{equation}
If $c^2/\Lambda \leq 10^{12}$ we get $\xi_* \leq 15$. On using
(\ref{lambdasol}) it is now possible to determine the evolution of
the Hubble parameters as functions of $\xi$. From Eqs. (\ref{hT2}) and
(\ref{hversusomega}) we get
\begin{eqnarray}
H^{2}(\xi) &\sim & \exp(-6 \xi) \left[H^{2}(0) +{2\over 3}c^{2}
\left(1-{3\over 4}\epsilon \right)\xi 
\right]\,\,\, , \\
h(\xi) & \sim & 3 \exp (-3 \xi) H(0)[1 -3 \epsilon]\,\,\, .
\end{eqnarray}

Late-time evolution is ruled by Eq. (\ref{lateevol}). Since
$\Omega'(\xi)<4$ it therefore follows that asymptotically $\Omega$
reaches a positive constant value $\Omega_\infty$, so that $h$ vanishes in
the large-$\xi$ limit as expected. In particular, for large $\xi$
Eq. (\ref{lateevol}) can be linearized, i.e.
\begin{equation}
\Omega''(\xi) \sim -3 \Omega'(\xi)\,\,\, ,
\end{equation}
and hence it is immediate to get the approximate solution
\begin{eqnarray}
H^2(\xi) &\sim&
\frac{\Lambda}{3}+\left(H^2(\xi_*)-\frac{\Lambda}{3} \right)
\exp(-6 (\xi-\xi_*)) \,\,\, , \\
h(\xi) &\sim& h(\xi_*) \exp(-3(\xi-\xi_*)) \,\,\, .
\end{eqnarray}

It should be stressed that the action functional studied in
the present Section or in Sec. II is not a low-energy limit of
string theory with a constant dilaton, since the starting point
remains a nonlocal action functional with $*$ product of fields.
Readers interested in string cosmology can be referred, for
example, to the work in Refs. \cite{[39],[40]}. In particular,
the work in Ref. \cite{[41]} has studied string cosmology with
a time-dependent antisymmetric tensor in a Bianchi I universe,
but as is stressed in Sec. IV of Ref. \cite{[41]}, since a pure
radiation plus dilaton solution has $\phi \rightarrow$ const, a
late-time isotropic radiation-dominated solution is a contracting
universe. By contrast, in our model, when the universe reaches
an isotropic stage it is still expanding.

\section{Concluding remarks}

Motivated by cosmology and noncommutative geometry,
we have investigated the effects of an antisymmetric tensor in
a Bianchi I early universe. What we do only holds at an 
intermediate stage where departure from ordinary spacetime
geometry can be appreciated, while lack of associativity of the
resulting $*$ product in curved spacetime is negligible \cite{[16]}.
Moreover, such a stage implies a departure from the models
with constant $\theta^{\mu \nu}$ which are relevant 
for string theory \cite{[21]}.
With this understanding, our original results are as follows.
\vskip 0.3cm
\noindent
(i) In the first model, where the antisymmetric tensor $\theta$
resulting from noncommutative geometry and a minimally coupled
scalar field driving inflation both occur, we have found by
numerical methods that a suitable range of the e-folding exists
such that $\theta$ prevails on the scalar field and the
longitudinal Hubble parameter prevails on $H_{\perp}$. This is
the anisotropic era, but for larger values of the e-folding the
antisymmetric tensor is damped and the scalar field dominates,
leading in turn to isotropization. The very existence of the
anisotropic era depends of course on the initial conditions chosen,
but it appears interesting to have shown explicitly how the
corresponding model can be built.
\vskip 0.3cm
\noindent
(ii) In the second model, again in a Bianchi I universe, gravity
with a cosmological constant $\Lambda$ is coupled to an
anti-symmetric tensor. The resulting nonlinear system of equations
for the Hubble parameters $H$ and $h$ has been solved by first
obtaining the differential equation (3.27) for the unknown
function $\Omega$, and then finding approximate solutions 
when the effects of $\theta_{\mu \nu}$ prevail upon $\Lambda$,
or the other way around. An accurate analytic description of the
process leading to an isotropic final state of the universe has
been therefore obtained.

An outstanding problem in sight is now the development of
cosmological perturbations' theory \cite{[38]}
within such a framework, with the
hope of obtaining quantitative information on the effect of $\theta$
on the formation of structure in the early universe.

\acknowledgments

We are indebted to Fedele Lizzi for enlightening conversations.
The work of G. Esposito has been partially supported by
PRIN 2002 ``Sintesi''; the work of G. Mangano and G. Miele
has been partially supported by COFIN 2002 
``Fisica Astroparticellare''.

\appendix
\section*{}
In this Appendix we consider the admissibility of the gauge condition
$\theta_{0i}= 0$ from the point of view of constraint analysis, and we
begin with the simplest model, with action functional
(as in Secs. II and III, the starting point is a nonlocal
action functional, which reduces to a local action by retaining
only quadratic terms in $\theta^{\mu \nu}$)
\begin{equation}
S=\int {\cal L}\, \sqrt{-g} \, d^{4}x \,\,\, ,
\label{(A1)}
\end{equation}
with Lagrangian density ${\cal L}
\equiv {1\over 12} H^{\mu \nu \rho} H_{\mu
\nu \rho}$ in a Bianchi I background, and field strength given by
cyclic permutations of $\nabla \theta$ terms, according to Eq.
(2.5). Only the effects of partial derivatives survive in
$H_{\mu \nu \rho}$, and hence we find, by virtue of (2.1),
\begin{equation}
{\cal L}={1\over 12}H^{\mu \nu \rho}H_{\mu \nu \rho}
={1\over 4}g^{ij}g^{kl}\theta_{ik,0}\theta_{jl,0} \,\,\, ,
\label{(A2)}
\end{equation}
since the assumption of spatial homogeneity implies
that $\theta^{\mu \nu}$ can only depend on $t$.
The term $g^{ij} \theta_{0i,0}\theta_{0j,0}$ is weighted with
coefficient $3-3=0$ and hence does not occur in
Eq. (\ref{(A2)}). The Lagrangian $L$ is obtained from $\cal L$ by
means of (recall that the lapse function $N$ is equal to $1$
in a Bianchi I background, and that $h$ denotes the determinant
of the induced three-metric)
\begin{equation}
L=\int {\cal L}  \, \sqrt{h} \, d^{3}x \,\,\, ,
\label{(A3)}
\end{equation}
and hence we find
\begin{equation}
\pi^{ij} \equiv {\delta L \over \delta \theta_{ij,0}}
={\sqrt{h}\over 2}g^{ir}g^{js}\theta_{rs,0} \; \; \; ,
\label{(A4)}
\end{equation}
\begin{equation}
\pi^{0i} \equiv {\delta L \over \delta \theta_{0i,0}}
\approx 0 \,\,\, .
\label{(A5)}
\end{equation}
Equation (\ref{(A5)}) deserves some comments: since the Lagrangian
is independent of $\theta_{0i,0}$, the momentum conjugate to
$\theta_{0i}$ vanishes. More precisely, the $\pi^{0i}$ can be seen
as $3$ primary constraints arising from the structure of the
Lagrangian; as such, their vanishing is only a weak equation
($\approx 0$), because $\pi^{0i}$ are well defined over the whole
phase space of the theory, and only vanish on the constraint
sub-manifold \cite{[42]}.

Our Lagrangian (\ref{(A3)}) reads
\begin{equation}
L=\int {1\over 2}\pi^{ij} \, \theta_{ij,0} \, d^{3}x \,\,\, ,
\label{(A6)}
\end{equation}
with corresponding canonical Hamiltonian
\begin{equation}
H_{c} \equiv \int \pi^{ij} \, \theta_{ij,0} \, d^{3}x-L
=\int {1\over 2}\, \pi^{ij} \, \theta_{ij,0} \, d^{3}x \,\,\, .
\label{(A7)}
\end{equation}
By virtue of the primary constraints (\ref{(A5)}), however, the
Hessian matrix is singular, and the time evolution is only well
defined when the effective Hamiltonian ${\widehat H}$ is
considered. The latter is given by $H_{c}$ plus a linear
combination of primary constraints, i.e. \cite{[42]}
\begin{equation}
{\widehat H}=\int \left({1\over 2}\pi^{ij}\theta_{ij,0} +\mu^{i}\pi_{0i}
\right)d^{3}x \,\,\, .
\label{(A8)}
\end{equation}
Moreover, we are still free to impose supplementary (more
frequently called `gauge') conditions, here chosen in the form
\begin{equation}
\theta_{0i} \approx 0 \,\,\, .
\label{(A9)}
\end{equation}
By doing so, we choose to regard the gauge conditions as
constraint equations, in much the same way as the Coulomb gauge
can be treated as a constraint equation in Maxwell theory \cite{[43]}.
We are therefore working with an extended Hamiltonian
\begin{equation}
H_{e} \equiv \int \left({1\over 2}\pi^{ij}\theta_{ij,0}
+\mu^{i}\pi_{0i}+\lambda^{i}\theta_{0i}
\right)d^{3}x \,\,\, ,
\label{(A10)}
\end{equation}
where the Lagrange multipliers $\mu^{i},\lambda^{i}$ can
be evaluated by requiring preservation in time of the primary
constraints $\pi_{0i}$ and gauge constraints $\theta_{0i}$.
For this purpose, note first that Eq. (\ref{(A4)}) yields
\begin{equation}
\theta_{ij,0}={2\over \sqrt{h}}\pi_{ij} \,\,\, ,
\label{(A11)}
\end{equation}
and hence (with $\rho \equiv {2\over \sqrt{h}}$)
\begin{equation}
H_{e}=\int \left({\rho \over 2}\pi^{ij}\pi_{ij}
+\mu^{i}\pi_{0i}+\lambda^{i}\theta_{0i}
\right)d^{3}x \,\,\, .
\label{(A12)}
\end{equation}
All our constraints are then trivially preserved, without
giving rise to further constraints, because
\begin{equation}
{d\over dt}\pi_{0i} \equiv
\left \{ \pi_{0i}({\vec x},t),H_{e} \right \}
\approx \int \lambda^{j}({\vec y},t)
\left \{ \pi_{0i}({\vec x},t),\theta_{0j}({\vec y},t)
\right \}d^{3}y=-\lambda_{i}({\vec x},t) \,\,\, ,
\label{(A13)}
\end{equation}
\begin{equation}
{d\over dt}\theta_{0i} \equiv
\left \{ \theta_{0i}({\vec x},t),H_{e} \right \}
\approx \int \mu^{l}({\vec y},t)
\left \{ \theta_{0i}({\vec x},t),\pi_{0l}({\vec y},t)
\right \}d^{3}y=\mu_{i}({\vec x},t) \,\,\, .
\label{(A14)}
\end{equation}
Note that, by virtue of our gauge constraints (\ref{(A9)}), the set of
constraints has been turned into the second-class, a feature shared by all
field theories after a gauge condition has been imposed \cite{[43]},
\cite{[44]}.

For the model studied in Sec. III, the full Hamiltonian constraint (3.11),
when expressed in integral form,
is the sum of (A12) and of the gravitational contribution.
The latter is obtained from spatial integration of
$$
(16\pi G) G_{ijkl}\tilde\pi^{ij}\tilde\pi^{kl} +\frac{\sqrt {h} \,
\,  ^{(3)}R }{16\pi G} +\Lambda \frac{\sqrt h}{8\pi G} \,\,\, ,
$$
where
$G_{ijkl}\equiv \frac{1}{2\sqrt h}(h_{ik}h_{jl}
+h_{il}h_{jk}-h_{ij}h_{kl})$
is the DeWitt supermetric on the space of Riemannian geometries on
$\Sigma$ \cite{[45]}, $\tilde\pi^{ij}$ is the momentum
conjugate to the induced three-metric and ${ }^{(3)}R$ is the
three-dimensional scalar curvature (our sign for such a
curvature is opposite to the one of Ref. \cite{[45]}).


\begin{references}
\bibitem {[1]}
J. Polchinski, \emph{
String Theory. Vol.1: An Introduction to the Bosonic String}
(Cambridge University Press, Cambridge, 1998).
\bibitem {[2]}
J. Polchinski, \emph{ String Theory. Vol.2: Superstring Theory and
Beyond} (Cambridge University Press, Cambridge, 1998).
\bibitem {[3]}
A. Connes, M.R. Douglas and A. Schwarz, J. High Energy Phys. {\bf
9802}, 003 (1998).
\bibitem {[4]}
A. Connes,
\emph{Noncommutative Geometry} (Academic Press, New York, 1994).
\bibitem {[5]}
A. Connes, Compt. Rend. Acad. Sci.(Ser.I Math.),
A {\bf 290}, 599 (1980) [hep-th 0101093].
\bibitem {[6]}
G. Landi,
\emph{An Introduction to Noncommutative Spaces and their Geometries}
(Springer, Berlin, 1997).
\bibitem {[7]}
J. Madore,
\emph{An Introduction to Noncommutative Geometry and its Physical
Applications} (Cambridge University Press, Cambridge, 1999).
\bibitem {[8]}
J.M. Gracia-Bond\`{\i}a, J.C. Varilly and H. Figueroa,
\emph{Elements of Noncommutative Geometry}
(Birkh\"{a}user, Boston, 2001).
\bibitem {[9]}
A. Kempf, Phys. Rev. D {\bf 63}, 083514 (2001).
\bibitem {[10]}
J. Martin and R. Brandenberger, Phys. Rev. D {\bf 63},
123501 (2001).
\bibitem {[11]}
A. Kempf and J.C. Niemeyer, Phys. Rev. D {\bf 64},
103501 (2001).
\bibitem {[12]}
F. Lizzi, G. Mangano, G. Miele and G. Sparano, Int. J. Mod. Phys.
A {\bf 11}, 2907 (1996).
\bibitem {[13]}
F. Lizzi, G. Mangano, G. Miele and G. Sparano, Mod. Phys. Lett.
A {\bf 11}, 2561 (1996).
\bibitem {[14]}
F. Lizzi, G. Mangano, G. Miele and G. Sparano, Phys. Rev. D
{\bf 55}, 6357 (1997).
\bibitem {[15]}
F. Lizzi, G. Mangano and G. Miele, Mod. Phys. Lett. A {\bf 16},
1 (2001).
\bibitem {[16]}
F. Lizzi, G. Mangano, G. Miele and M. Peloso, JHEP
{\bf 06}, 049 (2002).
\bibitem {[17]}
F.C. Adams, J.R. Bond, K. Freese, J.A. Frieman and A.V. Olinto,
Phys. Rev. D {\bf 47}, 426 (1993).
\bibitem {[18]}
J.D. Barrow and A.R. Liddle, Phys. Rev. D {\bf 47}, 5219 (1993).
\bibitem {[19]}
J. Garcia--Bellido and A.R. Liddle, Phys. Rev. D {\bf 55},
4603 (1997).
\bibitem {[20]}
Q.G. Huang and M. Li, ``CMB Power Spectrum from Noncommutative
Spacetime'' (hep-th/0304203).
\bibitem {[21]}
N. Seiberg and E. Witten, JHEP {\bf 09}, 032 (1999).
\bibitem {[22]}
J.M. Bardeen, Phys. Rev. D {\bf 22}, 1882 (1980).
\bibitem {[23]}
M.J. Duff in Supergravity 1981,
eds. Ferrara S. and J.G. Taylor
(Cambridge University Press, Cambridge, 1982).
\bibitem {[24]}
P.G.O. Freund and R.I. Nepomechie, \emph{ Unified Geometry of
Antisymmetric Tensor Gauge Field and Gravity}
(Nat. Sci. Found. 1981).
\bibitem {[25]}
C. Pathinayake, A. Vilenkin and B. Allen, Phys. Rev. D
{\bf 37}, 2872 (1988).
\bibitem {[26]}
R.K. Kaul, Phys. Rev. D
{\bf 18}, 1127 (1978).
\bibitem {[27]}
C.R. Hagen, Phys. Rev. D
{\bf 19}, 2367 (1979).
\bibitem {[28]}
E. Sezgin and P. van Nieuwenhuizen, Phys. Rev. D
{\bf 22}, 1127 (1980).
\bibitem {[29]}
W. Mecklenburg and L. Mizrachi, Phys. Rev. D
{\bf 29}, 1709 (1984).
\bibitem {[30]}
S. Deser and E. Witten,
Nucl. Phys. {\bf B178}, 491 (1981).
\bibitem {[31]}
A.A. Slavnov and S.A. Frolov,
\emph{ Quantization of Interacting Antisymmetric Tensor Field}
DFPD 13/87 (1987).
\bibitem {[32]} S.P. de Alwis, M.T. Grisaru and L. Mezincescu, \emph{
Quantization and Unitarity in Antisymmetric Tensor Gauge Theories}
BRX TH-235 (1988).
\bibitem {[33]}
N.Yu. Obukhov, Phys. Lett.
{\bf 109B}, 195 (1982).
\bibitem {[34]}
Z. Tokuoka, Phys. Lett.
{\bf A87}, 215 (1982).
\bibitem {[35]}
I. Bena, Phys. Rev. D
{\bf 62}, 127901 (2000).
\bibitem {[36]}
S. Deguchi, T. Mukai and T. Nakajima,
hep-th/9804070.
\bibitem {[37]}
M. Abud, J.P. Ader and L. Cappiello, Nuovo Cim. A {\bf 105}, 1507
(1992).
\bibitem {[38]}
V.F. Mukhanov, H.A. Feldman and R.H. Brandenberger, Phys. Rep.
{\bf 215}, 203 (1992).
\bibitem {[39]}
J.E. Lidsey, D. Wands and E.J. Copeland, Phys. Rep. 
{\bf 337}, 343 (2000).
\bibitem {[40]}
R. Easther, K.I. Maeda and D. Wands, Phys. Rev. D {\bf 53},
4247 (1996).
\bibitem {[41]}
E.J. Copeland, A. Lahiri and D. Wands, Phys. Rev. D {\bf 51},
1569 (1995).
\bibitem {[42]} 
P.A.M. Dirac, {\em Lectures on Quantum Mechanics}
(Dover, New York, 2001).
\bibitem {[43]}
A. Hanson, T. Regge and C. Teitelboim,
\emph{Constrained Hamiltonian Systems},
Contributi del Centro Linceo Interdisciplinare di
Scienze Matematiche e loro Applicazioni, n. 22
(Accademia Nazionale dei Lincei, Roma, 1976).
\bibitem {[44]}
G. Esposito, \emph{ Quantum Gravity, Quantum Cosmology and
Lorentzian Geometries}, Lecture Notes in Physics Vol. m12
(Springer, Berlin, 1994).
\bibitem {[45]}
B.S. DeWitt, Phys. Rev. {\bf 160}, 1113 (1967).
\end{references}
\end{document}